\documentclass[graybox]{svmult}

\usepackage{mathptmx}       
\usepackage{helvet}         
\usepackage{courier}        
\usepackage{type1cm}        
\usepackage{makeidx}         
\usepackage{graphicx}        
\usepackage{multicol}        
\usepackage[bottom]{footmisc}

\def \d {{\rm d}}

\makeindex             

\begin{document}

\title*{Connection between horizons and algebraic type}
\author{Otakar Sv\'{\i}tek}
\institute{Otakar Sv\'{\i}tek \at Institute of Theoretical Physics, Faculty of Mathematics and Physics, Charles University in Prague, V Hole\v{s}ovi\v{c}k\'{a}ch 2, 180 00 Praha 8, Czech Republic , \email{ota@matfyz.cz}}
%
%
\maketitle

\abstract{We study connections between both event and quasilocal horizons and the algebraic type of the Weyl tensor. The relation regarding spacelike future outer trapping horizon is analysed in four dimensions using double-null foliation.}

\section{Introduction}
\label{sec:1}

We would like to, at least partially, understand how does the presence of some form of horizon restrict the possible algebraic types on it. Since algebraic type of a tensor is determined locally we need to characterize the horizon without employing global notions. We will concentrate on the Weyl tensor and Petrov types derived from it.

Event horizon is a global characteristic and the full spacetime evolution is necessary in order to localize it. In many situations this is not desirable or even attainable and therefore, over the past years different quasi-local characterizations of black hole boundary were developed. The most important ones being apparent horizon \cite{hawking-ellis}, trapping horizon \cite{hayward} and isolated or dynamical horizon \cite{ashtekar}. The basic {\it local} condition in the above mentioned horizon definitions is effectively the same: these horizons are sliced by marginally trapped surfaces with vanishing expansion of outgoing (ingoing) null congruence orthogonal to the surface. We adopt the so called spacelike future outer trapping horizon (SFOTH) which merges the properties of trapping and dynamical horizons.

Since event horizon in a static spacetime with a well-behaved matter is a Killing horizon one can use the local condition on stationary Killing field in such a situation avoiding the global nature of event horizon. This case was investigated by Pravda and Zaslavskii \cite{pravda} and we summarize their results in the next section. In the third section the relation of SFOTH to the algebraic type is derived.

\section{Killing horizons}
\label{sec:2}

Pravda and Zaslavskii \cite{pravda} studied curvature scalars in a general static spacetime possessing Killing horizon (generalizing previous results of \cite{medved} on high degree of symmetry of the Einstein tensor to the non-extremal case). They assumed regularity of all polynomial invariants of the Riemann tensor on a horizon and used two naturally preferred frames for calculations - the static observer and the freely falling observer frames. Note that the static frame is singular on the null horizon.

Assuming 1+1+2 decomposition and using the Gauss normal coordinates the metric takes the following form
\begin{equation}
\d s^2=-M^{2}\d t^2+\d n^2+\gamma_{ab}\d x^{a}\d x^{b}
\end{equation}
The stationary Killing field is $\xi^{\mu}=(1,0,0,0)$ with $M^{2}\equiv\xi^{\mu}\xi_{\mu}=0$ on the horizon. The tetrad adapted to the static observer's four-velocity and the Gaussian normal direction has the form
\begin{equation}\label{SO-frame}
l^{\mu}=\frac{1}{\sqrt{2}}({\textstyle \frac{1}{M}},1,0,0),\ n^{\mu}=\frac{1}{\sqrt{2}}({\textstyle \frac{1}{M}},-1,0,0),\ m^{\mu}=(0,0,m^{a})
\end{equation}
which immediately implies $\Psi_{4}=\bar{\Psi}_{0}$ and $\Psi_{3}=-\bar{\Psi}_{1}$. Next, one can express the Weyl tensor, the Riemann tensor etc. using the above decomposition in terms of 2-metric $\gamma_{ab}$, extrinsic curvature $K_{ab}$, lapse $M$ and their derivatives. Upon projecting the Weyl tensor onto the tetrad and taking the horizon limit $M\to 0$ one gets the Weyl scalars on the horizon. Petrov type is then determined based on curvature invariants $I,J$ and coefficients $K,L,N$
\begin{equation}
I=\Psi_{0}\Psi_{4}-4\Psi_{1}\Psi_{3}+3\Psi_{2}^{2},\ J={\rm det}\left(\begin{array}{ccc}
\Psi_{4} & \Psi_{3} & \Psi_{2}\\
\Psi_{3} & \Psi_{2} & \Psi_{1}\\
\Psi_{2} & \Psi_{1} & \Psi_{0}
\end{array}\right)
\end{equation}
\begin{equation}
	K=\Psi_{1}\Psi_{4}^{2} - 3\Psi_{4}\Psi_{3}\Psi_{2} + 2\Psi_{3}^{3},\ L=\Psi_{2}\Psi_{4} - \Psi_{3}^{2},\ N=12 L^{2} - \Psi_{4}^{2}I
\end{equation}
The resulting Petrov type is either D ($\Psi_{2}\neq 0$) or O ($\Psi_{2}=0$).

In the case of the freely falling observer the adapted tetrad is given by simple transformation from (\ref{SO-frame})
\begin{equation}
\hat{l}^{\mu}=zl^{\mu},\ \hat{n}^{\mu}=z^{-1}n^{\mu}
\end{equation}
where $z=exp(-\alpha)$, ${\rm cosh}\alpha= \frac{E}{M}$, with $E$ being an energy per unit mass for radially infalling geodesic. In this frame invariants $I,J$ do not change but the coefficients are modified
\begin{equation}
\hat{K}=z^{-3}K,\ \hat{L}=z^{-2}L,\ \hat{N}=z^{-4}N
\end{equation}
Since $z\to 0$ on the horizon the coefficients $\hat{K},\hat{L},\hat{N}$ can be nonzero in the limit (unlike for static observer). The Petrov type is either II ($\Psi_{2}\neq 0$) or III ($\Psi_{2}=0$) here. Due to singular nature of the static frame on the horizon these results are more physically relevant.

\section{Quasilocal horizons}
As mentioned in the Introduction in the general dynamical situation we use the SFOTH - spacelike future outer trapping horizon - which has the following properties:
\begin{enumerate}
\item spacelike submanifold foliated by closed 2-surfaces with null normal fields $l,n$
\item expansion $\Theta_{l}=0$  (marginal)
\item expansion $\Theta_{n}<0$  (future)
\item ${\cal{L}}_{n}\Theta_{l}<0$ (outer)
\end{enumerate}
We employ a double-null foliation developed by Hayward \cite{hayward} (mainly for the characteristic initial value problem and the trapping horizon definition) and adapted by Brady and Chambers \cite{brady} to study a nonlinear stability of Kerr-type Cauchy horizons.

The procedure is based on a local foliation by closed orientable 2-surfaces $S$ with smooth embedding $\phi : S \times [0,U) \times [0,V) \to {\cal{M}}$ and induced spatial metric $h_{ab}$ on $S$ with corresponding covariant derivative $D_{a}$. Null vectors $l^{\mu}$, $n^{\mu}$ are normal to $S$ and there is a spatial two vector $s^{a}$ called shift (encoding freedom in identifying points on subsequent surfaces). Evolution of the induced metric is described using Lie derivatives along $l$ and $n$
\begin{eqnarray}
\Sigma_{ab}={\cal L}_{l}h_{ab}\, &,&\ \tilde{\Sigma}_{ab}={\cal L}_{n}h_{ab}\\
\theta={\textstyle \frac{1}{2}}h^{ab}\Sigma_{ab}\, &,&\ \tilde{\theta}={\textstyle \frac{1}{2}}h^{ab}\tilde{\Sigma}_{ab}\\
\sigma_{ab}=\Sigma_{ab}-\theta h_{ab}\, &,&\ \tilde{\sigma}_{ab}=\tilde{\Sigma}_{ab}-\tilde{\theta} h_{ab}\\
\omega_{a}&=&{\textstyle \frac{1}{2}}h_{ab}{\cal L}_{l}s^{b}
\end{eqnarray}
$\theta, \tilde{\theta}$ being expansions, $\sigma_{ab}, \tilde{\sigma}_{ab}$ shears and $\omega_{a}$ anholonomicity (related to normal fundamental form). We assume normalized null vectors thus having zero inaffinities.

From vacuum Einstein equations and contracted Bianchi identities one obtains
\begin{eqnarray}
{\cal{L}}_{l}\theta&=&-{\textstyle \frac{1}{2}}\theta^{2} - {\textstyle \frac{1}{4}}\sigma_{ab}\sigma^{ab}\\
{\cal{L}}_{l}h&=&\theta h\\
{\cal{L}}_{l}\omega_{a}&=&-\theta\omega_{a}+{\textstyle \frac{1}{2}}D^{b}\sigma_{ab} -{\textstyle \frac{1}{2}}D_{a}\theta \label{dyn-eq-omega}\\
{\cal{L}}_{l}(h^{-1/2}h_{ab})&=&h^{-1/2}\sigma_{ab}\label{dyn-eq-h}
\end{eqnarray}

After expressing curvature tensors in the given frame we get the following Weyl scalars in vacuum
\begin{eqnarray}
4\Psi_{0}&=&(2{\cal L}_{l}\Sigma_{ab}-\Sigma_{am}h^{mn}\Sigma_{bn})m^{a}m^{b}\label{Psi0}\\
4\Psi_{1}&=&(2\omega^{m}\Sigma_{am}+4{\cal L}_{l}\omega_{a})m^{a}\\
4\Psi_{2}&=&(2{\cal L}_{n}\Sigma_{ab}-4D_{a}\omega_{b}-\Sigma_{am}h^{mn}\tilde{\Sigma}_{bn}-4\omega_{a}\omega_{b})m^{a}\bar{m}^{b}\\
4\Psi_{3}&=&-(2\omega^{m}\tilde{\Sigma}_{am}+4{\cal L}_{n}\omega_{a})\bar{m}^{a}\\
4\Psi_{4}&=&(2{\cal L}_{n}\tilde{\Sigma}_{ab}-\tilde{\Sigma}_{am}h^{mn}\tilde{\Sigma}_{bn})\bar{m}^{a}\bar{m}^{b}
\end{eqnarray}

Next, we use the vanishing of expansion and further fixing of the spatial part of the frame for simplification. We evaluate the first term on the right-hand side of equation (\ref{Psi0}) noting that 
\begin{equation}
{\cal L}_{l}\Sigma_{ab}={\cal L}_{l}\sigma_{ab}+\theta\Sigma_{ab}+h_{ab}{\cal L}_{l}\theta
\end{equation}
Using the projection and the horizon condition we obtain
\begin{equation}
m^{a}m^{b}{\cal L}_{l}\Sigma_{ab}=m^{a}m^{b}{\cal L}_{l}\sigma_{ab}
\end{equation}
Assuming (see \cite{brady}) the Lie-propagated spatial part of the frame and equation (\ref{dyn-eq-h}) we arrive at 
\begin{equation}
0={\cal L}_{l}(h^{-1/2}h_{ab}m^{a}m^{b})=h^{-1/2}\sigma_{ab}m^{a}m^{b}
\end{equation}
Next, we may assume that $\omega_{a}=0$ initially on $S$ and would like to have ${\cal L}_{l}\omega_{a}=0$ as well. Indeed, the first and the last terms of equation (\ref{dyn-eq-omega}) vanish on the horizon and by further locally fixing the spatial part of the frame we obtain $m^{a}D^{b}\sigma_{ab}=0$. In a similar way, one can show that $\Sigma_{am}h^{mn}\Sigma_{bn}m^{a}m^{b}=0$ on the horizon.

Then $\Psi_{0}=0$ and $\Psi_{1}=0$. Assuming regularity of $\Psi_{\{2,3,4\}}$ we have $I^{3}=27J^{2}$ and therefore Petrov type II. Clearly the spacetime is generically type I away from the horizon.

In the future, we would like to check whether stronger statements are possible (with additional assumptions), generalize the results to well-behaved matter fields and nonzero cosmological constant. Also, we would like to extend the analysis to other important tensors (Ricci etc.).

\begin{acknowledgement}
This work was supported by grant GACR 202/09/0772 and project UNCE 204020/2012.
\end{acknowledgement}

\end{document}